 \documentclass[final,3p,times,onecolumn]{elsarticle}

 \usepackage{graphics}
 \usepackage{graphicx}

\usepackage{amssymb}
 \usepackage{amsmath}

\journal{Journal of Computational Physics}

\begin{document}

\begin{frontmatter}



\title{On the velocity space discretization for the Vlasov-Poisson system: comparison between Hermite spectral and Particle-in-Cell methods.\\
Part 2: fully-implicit scheme}


\author[1]{E. Camporeale}
\author[1]{G.~L. Delzanno}
\author[2]{B.~K. Bergen}
\author[1]{J.~D. Moulton}

\address[1]{T-5 Applied Mathematics and Plasma Physics, Los Alamos National Laboratory, 87545 Los Alamos, NM, USA.}
\address[2]{CCS-7 Applied Computer Science, Los Alamos National Laboratory, 87545 Los Alamos, NM, USA.}
\begin{abstract}
We describe a spectral method for the numerical solution of the Vlasov-Poisson system where the velocity space is decomposed by means of an Hermite basis, and the
configuration space is discretized via a Fourier decomposition.
The novelty of our approach is an implicit time discretization that allows exact conservation of charge, momentum and energy.
The computational efficiency and the cost-effectiveness of this method are compared to the fully-implicit PIC method recently 
introduced by \citet{markidis11} and \citet{chen11}. The following examples are discussed: Langmuir wave, Landau damping, ion-acoustic wave, two-stream instability.
The Fourier-Hermite spectral method can achieve solutions that are several orders of magnitude more accurate at a fraction of the cost with respect to PIC.
This paper concludes the study presented in \citet{camporeale13a} where the same method has been described for a semi-implicit time discretization, and was compared against an explicit PIC.
\end{abstract}


\end{frontmatter}


\section{Introduction}\label{Intro}
The Vlasov-Poisson system describes the dynamics of an electrostatic collisionless plasma. The three main numerical approaches to solve it 
are: Particle-in-Cell (PIC), Eulerian and semi-Lagrangian Vlasov solvers, and spectral Vlasov solvers. 
PIC simulations probably represent the most popular method in the plasma physics community. This might be due to
its algorithmic simplicity, the relative ease of its parallelization, and the recent impressive advances in available computer power.
The main idea of PIC consists in using super-particles to sample the distribution function in velocity space, and to mediate particles interactions through
a computational grid where the electromagnetic field is calculated \citep{birdsall_book,hockney_book}.
The well-known shortcoming of PIC codes is the intrinsic noise associated with the discrete nature of super-particles: as
typical of Monte Carlo methods, the numerical error scales as $\sim N_p^{-1/2}$, while the computing time scales as $\sim N_p$ (with $N_p$ the number of particles) \citep{verboncoeur05}.
For this reason alternative methods for solving the Vlasov equation have been studied in the last three decades (see, e.g., \citep{shoucri08}), in addition to PIC methods 
that can reduce particle noise (such as the $\delta$f method or methods based on remapping the distribution function, see for instance \citep{chehab05, wang11, deng14}).\\
In the companion paper (Part 1) \citep{camporeale13a}, we have discussed a spectral discretization of the velocity space for the Vlasov equation by means of an orthogonal
Hermite basis. This approach has been studied in the past by \citet{grad49, engelmann63, grant67a,armstrong70,holloway96,schumer98,lebourdiec06,camporeale06,siminos11}.\\
In Part 1, we have: 1) introduced a semi-implicit time discretization which is more stable than the explicit discretization discussed in Ref. \citep{schumer98}; 
2) discussed the effect of an artificial collisional operator and introduced an operator which is charge, momentum, and energy conserving;
3) performed a qualitatively comparison between the performances of the Fourier-Hermite (FH) and explicit PIC methods in terms of computational efficiency and efficacy. 
The comparisons were carried out with the standard PIC algorithm, without any denoising applied, in part in an attempt to quantify the real limitations of each algorithm 
in its most basic form and in part because the FH method is less mature than PIC and optimization methods to accelerate the convergence of the 
Hermite series are in practice still unavailable. On standard benchmark cases typical of the PIC community (involving mostly near-equilibrium Maxwellian plasmas), 
we have shown that the FH method is orders of magnitude more efficient than PIC, and that the standard PIC method is essentially 
unable to capture low amplitude nonlinear interactions such as echo phenomena, despite
using a very large number of particles per cell (up to $2\cdot 10^6$ were used in \citep{camporeale13a}, much larger than what is normally used).\\
The results of Part 1 were limited to the comparison between an explicit PIC and a semi-implicit FH method.
The goal of this paper is to extend the study presented in Part 1 to a fully-implicit time discretizations.
Explicit PIC has often been criticized due to its stringent requirements on the choice of the time step and grid size, for numerical stability reasons.
In fact, an explicit PIC code requires the resolution of the smallest time scale and the shortest
spatial scale, even when the physics of interest only involves larger time/spatial scales.
Of course, this generally translates into the requirement for large computational resources.
Historically, the search for more efficient PIC schemes based on implicit time discretization dates back to the 80's, when
the implicit moment \citep{mason81,brackbill82,markidis09} and the direct implicit \citep{cohen82,langdon83} methods were introduced.
Both methods rely on a linearization of the equations for the electromagnetic fields, and, thus, they should be more properly
regarded as \textit{semi-implicit} methods.
Using these techniques, the numerical stability is greatly improved (with respect to explicit PIC), but energy is still not conserved exactly and, at each time step,
there is a small inconsistency between the charge and current densities calculated from the particles and the one that is used for advancing the fields. 
Some authors have suggested that such limitations are responsible for the accumulation of numerical errors that preclude semi-implicit PIC simulations
to run for long time intervals \citep{chen11,chen13}. Recently, however,
\citet{markidis11} and \citet{chen11} have formulated and successfully implemented a fully-implicit, one-dimensional PIC code. The main feature that 
characterizes fully-implicit PIC is that particles and fields are advanced simultaneously through a Jacobian-Free Newton-Krylov (JFNK) solver, and converged
nonlinearly within a certain tolerance.
Moreover, by using the so-called particle enslavement, the nonlinear solver converges on a residual that does not contains particle positions and velocities. 
In the fully-implicit PIC method, energy is conserved within the level imposed by 
the nonlinear tolerance (i.e. almost exactly). \citet{chen11} have shown that by implementing a sub-stepping procedure that makes particles stop every time they cross a cell boundary, the charge is also conserved
exactly. On the other hand, momentum is not conserved and must be monitored throughout the simulation (in contrast to semi-implicit PIC that exactly conserves the momentum, but not the energy).\\ 
In this work we devise a fully-implicit FH method. One of the important features of the method is that it simultaneously conserves charge, momentum and energy. 
In this regard, it is superior to any existing PIC method, for which it is impossible to conserve both momentum and energy. 
This paper concludes the study initiated in Part 1, by comparing the PIC and the FH method, both in their fully-implicit version.
Of course, the issue related to numerical noise still holds for the fully-implicit PIC, and for this reason we will show that the conclusions of this work 
are consistent with the results presented in Part 1: at least on the benchmark tests considered, the FH method is orders of magnitude more efficient than the PIC method. \\
The paper is organized as follows. In Section 2, we introduce the Fourier-Hermite expansion of the Vlasov-Poisson system in one dimension (1D). We use a fully-implicit discretization
of the equation based on the Crank-Nicolson scheme, which leads to a non-linear system of equations that is solved numerically by means of a JFNK solver.
We also show that the fully-implicit discretization ensures exact charge, momentum and energy conservation.
Section 3 presents the comparison between FH and PIC for four standard cases: Langmuir wave, linear Landau damping, two-stream instability, and ion acoustic wave.
We present conclusions in Section 4.

\section{Numerical Method}\label{hermite}
We study the Vlasov-Poisson system in the 2-dimensional phase space $(x,v)$, where $x$ denotes position and $v$ velocity.  
The phase space is assumed to be periodic in $x$.
In order to describe the method we specialize to the case of a plasma consisting of electrons and singly charged ions. 
The quantities are normalized as follows. Time is normalized to the electron plasma frequency $\omega_{pe}$, velocities to
the electron thermal velocity $v_{te}=\sqrt{kT_e/m_e}$, lengths to the electron Debye length $\lambda_D$, 
the electric field to $\frac{m_e}{e}v_{te}\omega_{pe}$ ($k$ is the Boltzmann's constant, $T_e$ is the electron temperature,
$m_e$ is the electron mass and $e$
is the elementary charge), densities to a reference density $n_0$.
The Vlasov equation for species $s$ reads:
\begin{equation}
 \frac{\partial f_s}{\partial t} + v\frac{\partial f_s}{\partial x} + \frac{q_sm_e}{em_s} E\frac{\partial f_s}{\partial v}=0\label{vlasov}
\end{equation}
Here $f_s$ is the particle distribution function, $q_s$ and $m_s$ are the charge and mass of the particles of species $s$ ($s=e,i$ for electrons and ions, respectively) and $E$ is the electric field.
The electric field is self-consistently calculated through the Poisson equation:
\begin{equation}
\frac{\partial E}{\partial x}= \int^\infty_{-\infty} f_i dv-\int^\infty_{-\infty} f_e dv.\label{poisson}
\end{equation}
The Hermite basis $\Psi_n(\xi)$ and the Fourier basis $\Phi_k(x)$ are defined as:
\begin{eqnarray*}
 \Psi_n(\xi)&=&(\pi2^nn!)^{-1/2}H_n(\xi)e^{-\xi^2}\label{H_basis_1},\\
 \Phi_k(x)&=& \exp(2\pi ikx/L),
\end{eqnarray*}
where $\xi_s=(v-u_s)/\alpha_s$, $H_n$ is the $n$-th Hermite polynomial, $L$ is the box length in $x$, and $u_s$ and $\alpha_s$ are two free parameters that can be used to improve the convergence of the series \citep{camporeale06}.
We expand the distribution function $f_s$ as:
\begin{equation}
 f_s(x,v,t) = \sum_{n=0}^{N_H-1} \sum_{k=-N}^N C_{n,k}^s(t)\Psi_n(\xi_s)\Phi_k(x)
\end{equation}
Following the derivation outlined in Part 1, one obtains the following set of ODEs for the coefficients $C_{n,k}^s$:
\begin{equation}\label{vlasov_hf}
 \frac{dC^s_{n ,k}}{d t}  + \alpha_s\frac{2\pi ik}{L}\left(\sqrt{\frac{n+1}{2}}C^s_{n+1,k} + \sqrt{\frac{n}{2}} C^s_{n-1,k}+\frac{u_s}{\alpha_s}C^s_{n,k}\right) -\frac{q_sm_e}{em_s}
 \frac{\sqrt{2n}}{\alpha_s}\sum_{m=-N}^N E_{k-m} C_{n-1,m}^s=0.
\end{equation}
Equation (\ref{vlasov_hf}) is discretized in time with a second order accurate fully-implicit Crank-Nicolson scheme \citep{crank47}, which, for each species, reads (omitting subscript $s$):
\begin{eqnarray}
\hspace*{-10cm}\nonumber \frac{C_{n,k}^{t+1}-C_{n,k}^{t}}{\Delta t} &+&\alpha_s\frac{\pi ik}{L}\left[\sqrt{\frac{n+1}{2}}\left(C_{n+1,k}^{t+1}+C_{n+1,k}^t\right)+ \sqrt{\frac{n}{2}} \left(C_{n-1,k}^{t+1}+C_{n-1,k}^t\right)
+\frac{u_s}{\alpha}\left(C_{n,k}^{t+1}+C_{n,k}^{t}\right)\right]\\
&-&\frac{q_sm_e}{em_s} \frac{\sqrt{2n}}{4\alpha_s}\sum_{m=-N}^N \left(E_{k-m}^{t+1}+E_{k-m}^{t}\right) \left(C_{n-1,m}^{t+1}+C_{n-1,m}^{t}\right)=0,\label{vlasov_cn}
\end{eqnarray}
where subscript $t$ indicates the time step.
This is coupled to the Poisson equation in Fourier space:
\begin{eqnarray}\label{poisson_hf}
  E_k &=& \frac{iL}{2\pi k}\left(\alpha_eC_{0,k}^e-\alpha_i C_{0,k}^i\right)   \mbox{       for $k\neq0$,} \\
  E_0 &=& 0.
\end{eqnarray}
Note that the Fourier representation of the Poisson equation (\ref{poisson}) leaves the constant $E_0$ undefined, while imposes the constraint $\alpha_iC^i_{0,0}=\alpha_eC^e_{0,0}$
(which physically means that the plasma is neutral). 
However, the absence of an externally imposed electric field and the periodicity of the domain dictates that $E_0=0$.\\
In summary, the problem is reduced to the numerical solution of the set of nonlinear Eqs. (\ref{vlasov_cn}). We have employed a Jacobian-Free Newton-Krylov solver,
which algorithmically consists in defining Eq.(\ref{vlasov_cn}) as a residual that is iteratively minimized at each time step \citep{kelley_book}. 
\subsection{Conservation properties}
As discussed in Part 1, the FH discretization has the property of conserving charge and momentum exactly, for any time discretization.
The Crank-Nicolson scheme also allows exact conservation of the total energy.
The simultaneous conservation of charge, momentum and energy is an important property of this technique, especially when compared with PIC codes. 
Indeed, it is important to emphasize that no existing PIC code can simultaneously conserve energy and momentum. In practice, this means
that the non-conserved quantity must be monitored throughout the simulation in order to ensure that its error is somehow bounded.\\
The following proof of the conservation of energy for the FH scheme is, for simplicity, carried out for a single species. More general cases are straightforward, but with more cumbersome algebra.
The kinetic and potential energy $W_K$ and $W_E$ are defined as:
\begin{eqnarray}
 W_K &=& \frac{1}{2}\int_0^L \int_{-\infty}^\infty v^2f(x,v,t) dx dv = \frac{\alpha^3L}{4}(\sqrt{2}C_{2,0} + C_{0,0}) + \frac{u\alpha}{2}(\alpha\sqrt{2}C_{1,0} + u C_{0,0})\\
 W_E &=& \frac{L}{2}\sum_k |E_k|^2
\end{eqnarray}
The change of potential energy between time $t+1$ and time $t$ is
\begin{align*}
 \Delta W_E &= \frac{L}{2}\sum_k |E_k^{t+1}|^2-|E_k^t|^2=\frac{L^3\alpha^2}{8\pi^2}\sum_{k\neq 0}\frac{1}{k^2}(|C_{0,k}^{t+1}|^2-|C_{0,k}^t|^2)=\\
 &=\frac{L^3\alpha^2}{8\pi^2}\sum_{k\neq 0}\frac{1}{k^2}\left(|C_{0,k}^t-\frac{\Delta t\alpha\pi i k}{\sqrt{2}L}(C_{1,k}^{t+1}+C_{1,k}^t)|^2-|C_{0,k}^t|^2\right)=\\
 &=\sum_{k\neq 0} \frac{\alpha^4L\Delta t^2}{16}|C_{1,k}^{t+1}+C_{1,k}^t|^2\\
 &+ \frac{\sqrt{2}L^2\alpha^3\Delta t}{8\pi k}\left(\mathrm{Re}(C_{0,k}^t) \mathrm{Im}(C_{1,k}^{t+1} + C_{1,k}^t) - \mathrm{Im}(C_{0,k}^t) \mathrm{Re}(C_{1,k}^{t+1}+C_{1,k}^t)\right),
\end{align*}
where Eq.(\ref{vlasov_cn}) has been used. $\mathrm{Re}$ and $\mathrm{Im}$ indicate the real and imaginary parts of a complex quantity. Similarly, one can calculate the 
the change of kinetic energy between times  $t+1$ and time $t$, considering that $C_{0,0}^{t+1}=C_{0,0}^{t}$, and $C_{1,0}^{t+1}=C_{1,0}^{t}$.
\begin{align*}
 \Delta W_K &= \frac{\alpha^3L\sqrt{2}}{4}(C_{2,0}^{t+1}-C_{2,0}^t)=\frac{\alpha^3 L\sqrt{2}}{4}\left(-\frac{\Delta t}{2\alpha} \sum_{k\neq 0}(E_{-k}^{t+1}+E_{-k}^{t})(C_{1,k}^{t+1}+C_{1,k}^t) \right)=\\ 
 &=\frac{\Delta t \alpha^3 L^2i\sqrt{2}}{16\pi}\sum_{k\neq 0}\frac{1}{k}(C_{0,-k}^{t+1}+C_{0,-k}^t)(C_{1,k}^{t+1}+C_{1,k}^t)=\\
 &=\frac{\Delta t \alpha^3 L^2i\sqrt{2}}{16\pi}\sum_{k\neq 0}\frac{1}{k}\left(2C_{0,-k}^t + \frac{\Delta t \alpha \pi i k}{\sqrt{2}L}(C_{1,-k}^{t+1}+C_{1,-k}^t)  \right) (C_{1,k}^{t+1}+C_{1,k}^t)=\\
 & =-\sum_{k\neq 0}   \frac{\sqrt{2}L^2\alpha^3\Delta t}{8\pi k}\left(\mathrm{Re}(C_{0,k}^t) \mathrm{Im}(C_{1,k}^{t+1} + C_{1,k}^t) - \mathrm{Im}(C_{0,k}^t) \mathrm{Re}(C_{1,k}^{t+1}+C_{1,k}^t)\right) \\
 &+  \frac{\alpha^4L\Delta t^2}{16}|C_{1,k}^{t+1}+C_{1,k}^t|^2.
\end{align*}
Therefore, $\Delta W_E+\Delta W_K = 0$.

\subsection{Collisional term}
The role of a collisional term for the numerical solution of the Vlasov-Poisson equations for a collisionless plasma has been extensively
discussed in Part 1.
In short, the collisional term has the double role of limiting velocity filamentation and suppressing numerical instabilities.
Here we use the same collisional term employed in Part 1, namely the operator $\mathcal{C}$ acts
on the coefficient $C_{n,k}$ as:
\begin{equation}
 \mathcal{C}[C_{n,k}] =-\nu \frac{n(n-1)(n-2)}{(N_H-1)(N_H-2)(N_H-3)} C_{n,k}\label{coll_operator},
\end{equation}
where $\nu$ is the collisional rate applied to the last Hermite coefficients $C_{N_H-1,k}$.

We emphasize that this operator is constructed in such a way that the coefficients $C_{0,k}$, $C_{1,k}$, and $C_{2,k}$ are not modified with respect to
the exact collisionless case $\nu=0$. As such, charge, momentum, and energy are conserved even when collisions are applied.
The collision operator (\ref{coll_operator}) is applied on the right hand side on Eq. (\ref{vlasov_hf}).

\section{Results}\label{results}
Similarly to the procedure followed in Part 1 \citep{camporeale13a}, in this section we compare the performance of the Fourier-Hermite (FH) method with the PIC method, both implemented with fully-implicit time discretization.
For the fully-implicit PIC method, we follow the approach of \citet{chen11}, that nonlinearly solves the Ampere equation discretized in time with a Crank-Nicolson scheme.
The current density is self-consistently calculated from the particles.
The only difference between the fully-implicit PIC used in this paper and the one employed in \citet{chen11} is that we do not use a space filter (smoothing), for the following reason.
We have calculated the numerical plasma frequency from a PIC simulation of Langmuir wave in a cold plasma, and we have verified that its deviation from the theoretical
plasma frequency (equal to 1, in normalized units) is 4 times larger when using the binomial filter adopted in \citet{chen11}. 
The PIC code employs linear interpolation, usually referred to as 'Cloud-in-Cell' (CIC) \citep{birdsall_book}. 
Note that higher order interpolation schemes have been proposed, for instance, in \citep{lewis70,evstatiev13}.
In order to make the comparison as fair as possible, we keep the same number of grid points for FH and PIC.
Note that, by solving Ampere's instead of Poisson's equation, the fully-implicit PIC method does not require a space derivative operator. Hence, there is no equivalent of the
Fourier discretization in the PIC. Of course, the grid spacing introduces a numerical error in the current accumulation and interpolation routines.
Finally, although a preconditioned version of the fully-implicit PIC has been presented in \citet{chen13}, we use the unpreconditioned version here, so that both method are unpreconditioned.
Since the focus of this work is on the discretization in velocity space, i.e. the comparison
between the spectral Hermite method and the use of super-particles, all simulations are performed for the same choice of timestep and grid size.  
Following Part 1, the comparison is characterized by the following three metrics:
\begin{enumerate}
\item the error with respect to a 'reference' highly accurate solution as a function of CPU time and velocity discretization (number of 
Hermite modes $N_H$ for FH and number of particles per cell $N_{pcel}$ for PIC); 
\item the error with respect to the 'previous' less accurate solution; 
\item the efficacy defined as the inverse of the product of CPU time and error.
\end{enumerate}
The error used for all runs is calculated as the $L_1$ norm of the difference between two solutions, averaged in time.
The error in 2) is what is actually used by a user who is performing a convergence study to decide when the solution is accurate enough.
The efficacy is a useful indicator of the cost-effectiveness of an algorithm. It measures whether an additional cost in terms of CPU time is compensated by a gain in terms of accuracy.
Clearly, an algorithm performs well if the efficacy increases notably with increasing CPU time. In this regard, we 
repeat the argument made in Part 1, namely that the PIC algorithm, by construction, performs badly in terms of efficacy 
since the error scales as $N_{pcel}^{-1/2}$, while the computing time scales roughly linearly with $N_{pcel}$. 
Therefore, the efficacy scales as the inverse of the square root of the CPU time, 
i.e., it actually decreases with increasing
CPU time. Hence, from a pure cost-effectiveness point of view, it is \textit{never} advantageous to increase the number of particles in a PIC code to reduce the error. 
On the other hand, one is often forced to have a large number of particles such that 
the physical signal is above the noise level (see e.g. \citep{camporeale11}).\\
For all the cases discussed below we initialize the electrons (ions) with a Maxwellian distribution function with thermal velocity $\alpha_{e}$ ($\alpha_{i}$). 
For the Langmuir wave, the Landau damping and the two-stream instability tests
the ions constitute a fixed background, while for the study of the ion-acoustic wave they evolve.
The initial electric field is initialized as:
\begin{equation}
 E(x,t=0) = \frac{L}{2\pi}\varepsilon\sin(2\pi x/L),
\end{equation}
where $\varepsilon$ is the amplitude of the initial perturbation.
In Fourier space, such an initialization corresponds to:
\begin{equation}
 E_{-1}=E_{1}=-\frac{\varepsilon L}{4\pi}
\end{equation}
and the density is initialized consistently.
For all runs, the number of grid points (Fourier modes for FH) is equal to $2N+1=33$ and the timestep is $\Delta t=0.05$. 
We use the \texttt{nsoli} routine described in \citep{kelley_book} for the JFNK solver.
The Krylov solver is a non-preconditioned restarted GMRES. For all simulations reported in this paper, the Newton-Krylov absolute and relative tolerances are set to $10^{-8}$.
All the codes are written in MATLAB and run on an Intel Xeon 3.40 GHz Linux box.

\subsection{Langmuir wave}\label{langmuir}
The parameters are chosen as follows: $L=2\pi$, $\alpha_e=0.1\sqrt{2}$, $\varepsilon=0.01$, $\nu=0$.
Figure \ref{fig:langmuir} shows the comparison between FH and PIC (the same format is used for following figures).
The top panels show the error with respect to the 'reference' solution (red circles) and with respect to the 'previous' less refined solution (black circles).
The results for PIC are on the left, as a function of number of particles $N_{pcel}$, and the results for FH are on the right as a function of number of Hermite polynomials $N_H$.
The reference solutions are calculated with $N_{pcel}=102400$ and $N_H=100$, respectively for PIC and FH.
The PIC result recovers the theoretical scaling with $N_{pcel}^{-1/2}$ (black line).
For this case both PIC and FH reach a low error with a relatively low number of particles per cell and Hermite coefficients, respectively.
This was not the case with the explicit PIC, and thus one can infer that the conservation of energy is beneficial, in this case, to assure a fast convergence of the solution.
The FH method is advantageous with respect to PIC only if one aims to obtain extremely low errors (of the order of $10^{-12}$).
The two bottom panels of Figure \ref{fig:langmuir} show the error (left) and the efficacy (right) as a function of the 
CPU time. Black circles are for PIC and red circles are for FH. Here the error is with respect to the reference solution.
As a figure of merit, in order to reach an error of the order of $10^{-11}$, the FH method takes about 30 seconds,
while the PIC takes about 3300 seconds, i.e. more than 100 times longer. This is
reflected in computing the efficacy (right-bottom panel). As anticipated, the
PIC efficacy scales as the inverse of the square root of the CPU time (black line), i.e. it decreases with increasing CPU time.
In contrast, the FH efficacy increases by about 4 orders of magnitude when the CPU time increases 
by a factor of 4 (from 15 to 60 seconds).

\subsection{Landau damping}\label{landau}
\begin{figure}
\centering
\includegraphics[width=1\textwidth]{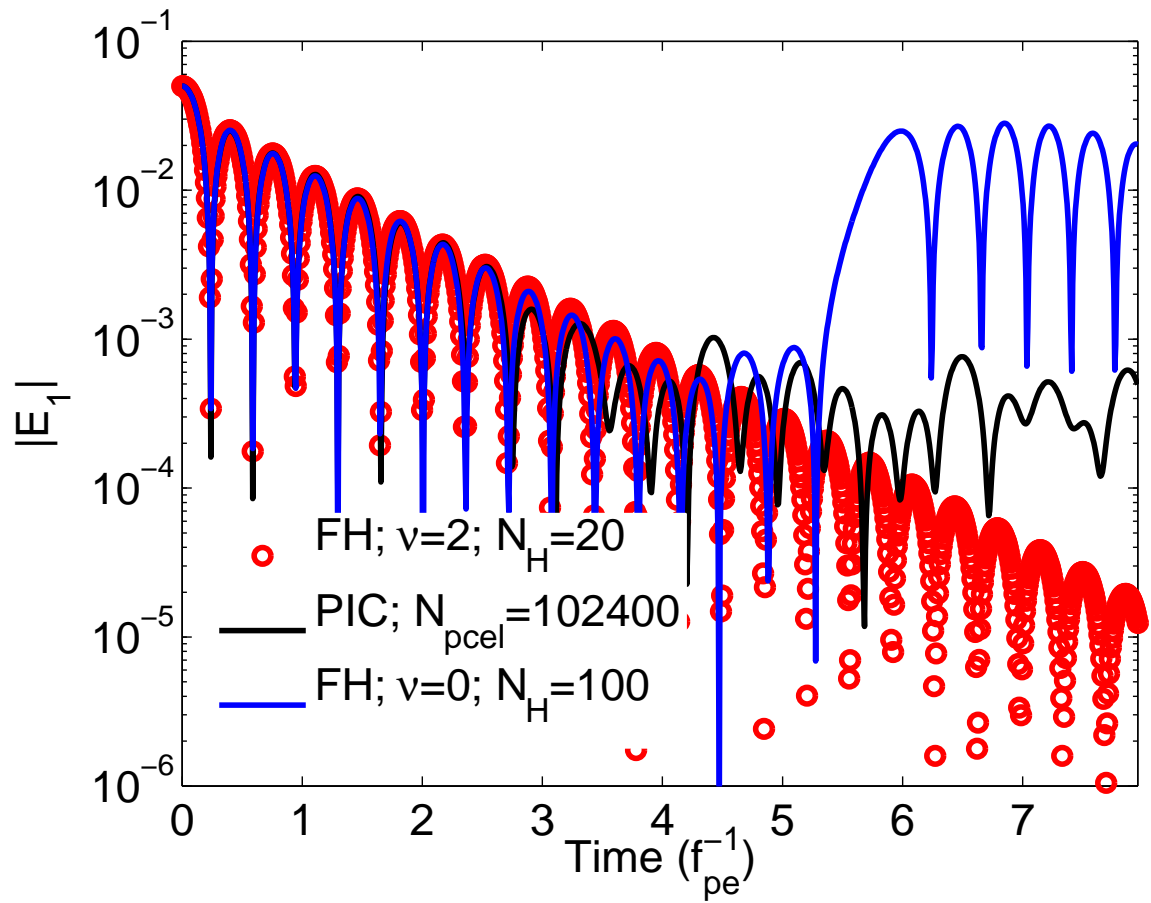}
\caption{Landau damping simulation. Black line: reference solution for PIC with $N_{pcel}=102400$; blue line: solution for FH with $\nu=0$, $N_H=100$; red circle: solution for FH with $\nu=2$; $N_H=20$}\label{fig:landau_nu}
\end{figure}

\begin{figure}
\centering
\includegraphics[width=1\textwidth]{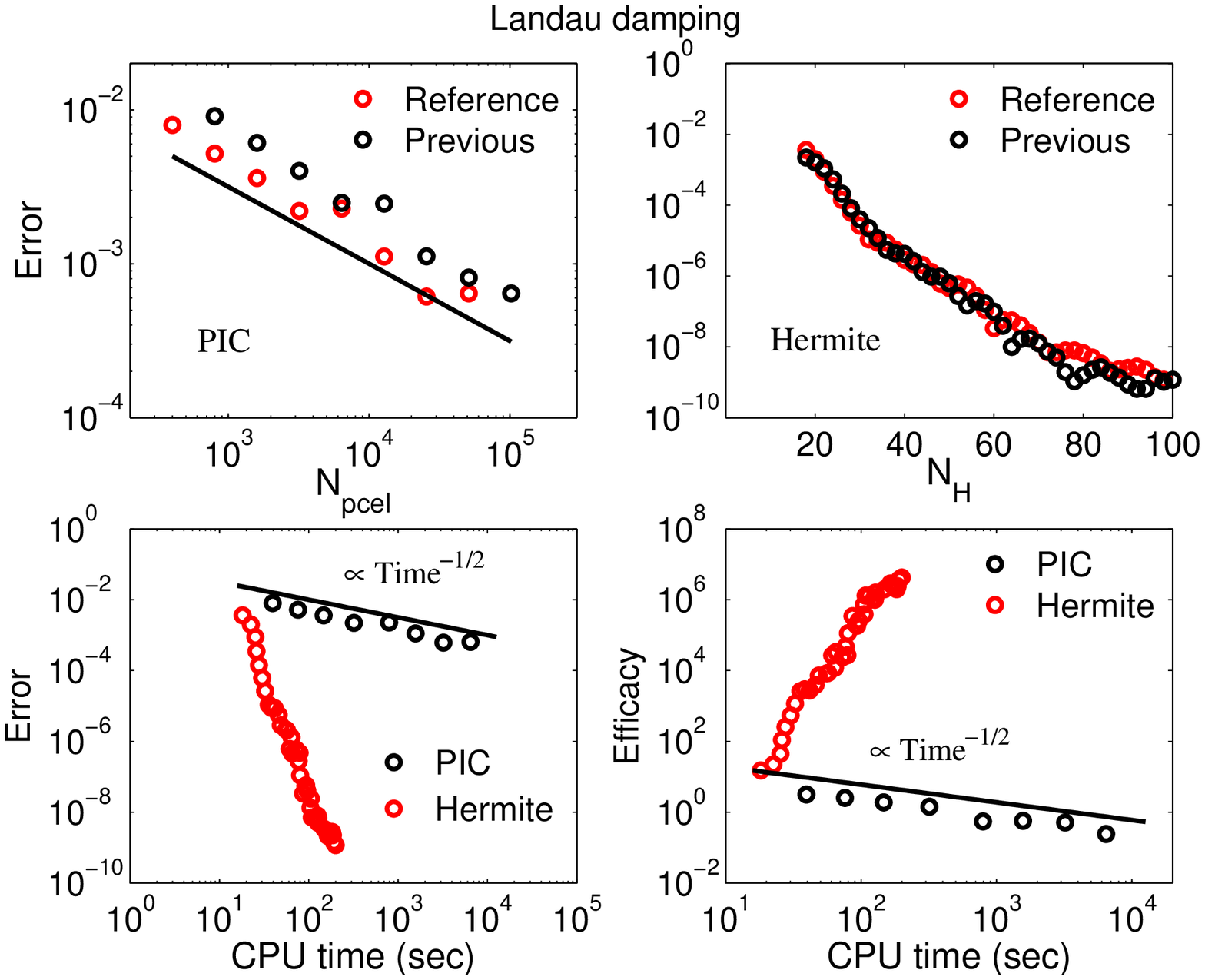}
\caption{Landau damping with parameters $L=4\pi$, $\alpha_e=\sqrt{2}$, $\varepsilon=0.05$, $\nu=0$. Top left: PIC simulation; error as a function of number of particles per cell $N_{pcel}$. Red and black circles represent the error
calculated with respect to a reference solution (with $N_{pcel}=102400$) and previous less accurate solution, respectively. The black solid line indicates the scaling $N_{pcel}^{-1/2}$.
Top right: Hermite simulation; error as a function of number of Hermite modes $N_H$. Red and black circles represent the error
calculated with respect to a reference solution (with $N_H=100$) and previous less accurate solution, respectively.
Bottom left: error as a function of CPU time (in seconds); black circles for PIC, red circles for Hermite. Bottom right: efficacy as a function of CPU time (in seconds); black circles for PIC, red circles for Hermite.
}\label{fig:landau}
\end{figure}

The Landau damping case is run with the following parameters: $L=4\pi$, $\alpha_e=\sqrt{2}$, $\varepsilon=0.05$, $\nu=0$.
The errors are averaged in the time window from $T=0$ to $T=3$. After $T=3$ the PIC reference solution (with $N_{pcel}=102400$) is affected by a low noise-to-signal ratio.
This is shown in Figure \ref{fig:landau_nu}. Here the black line is the PIC reference solution and the blue line is the result from FH with $N_H=100$.
Clearly both results do not suffer from recurrence/noise up to $T=3$. Additionally, we show with red circles
the result from FH when a collisionality $\nu=2$ is used, with $N_H=20$. Consistent with the results discussed in Part 1,
adding a collisional term allows the elimination of the recurrence problem for FH, even using a low number of Hermite polynomials.
Figure \ref{fig:landau} presents the comparison between FH and PIC with the same format of Figure
\ref{fig:langmuir}. Once again the correct scaling with the inverse of the number of particles per cell
is recovered for PIC. The difference in performance between the two methods is now 
even greater than for the Langmuir wave case.
For instance, an error equal to $6\cdot10^{-4}$ corresponds to a CPU time of 26 seconds for FH and about 
6400 seconds for the PIC, with a ratio between the two times approximately equal to 250.
Conversely a simulation that takes 200 seconds achieves errors equal to $10^{-3}$ and $10^{-9}$ for PIC and 
FH, respectively, i.e. FH is 6 orders of magnitude more accurate for equal CPU time.
On this test, we have also tried the space filtering (smoothing) described in \citep{chen11}, and verified that the results do not change qualitatively.

\subsection{Ion-acoustic wave}\label{ionacoustic}
\begin{figure}
\centering
\includegraphics[width=1\textwidth]{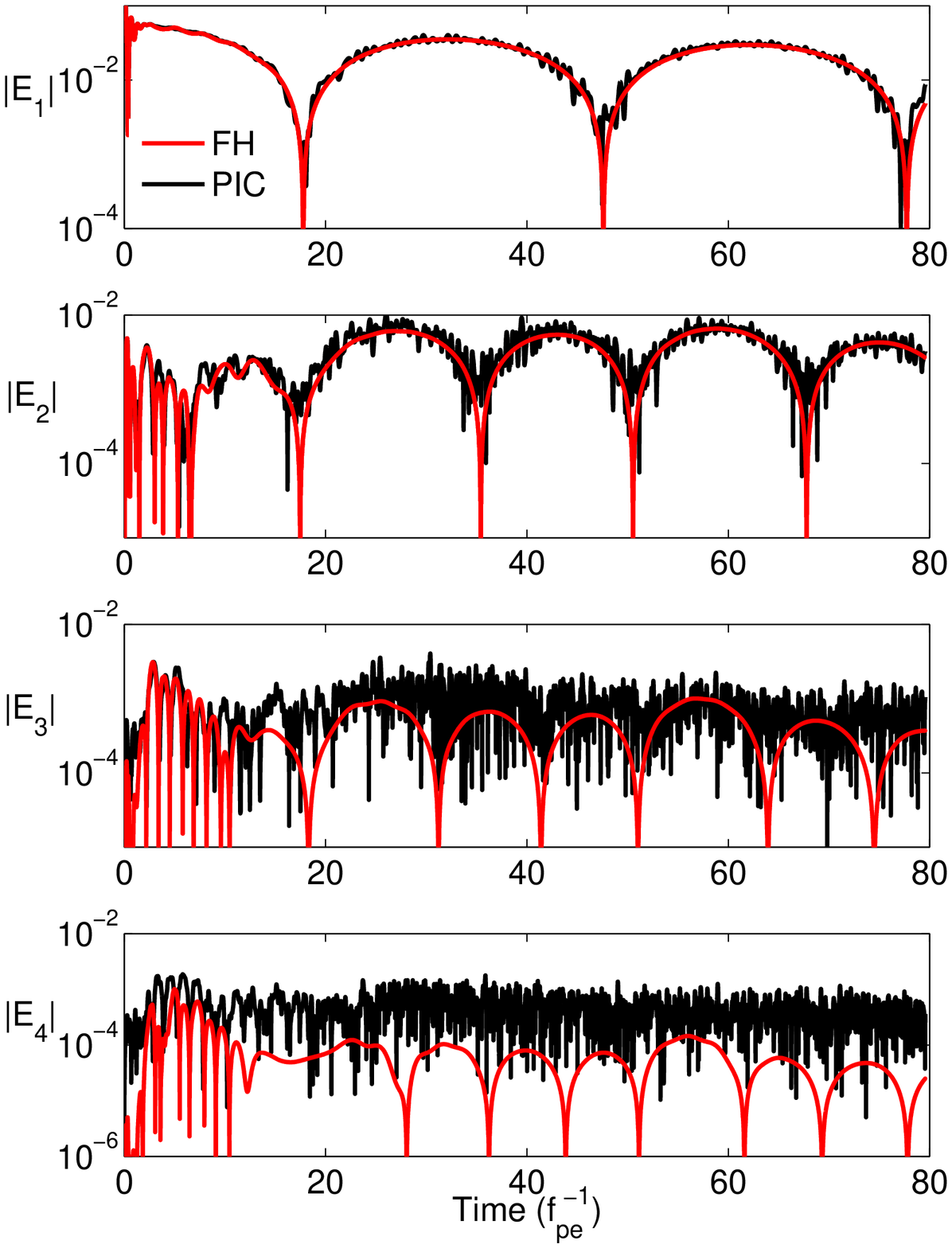}
\caption{Ion-acoustic wave simulation with parameters $L=10$, $\alpha_e=\sqrt{2}$, $\alpha_i=0.0074\alpha_e$, $\varepsilon=0.2$, $\nu=1$. The mass ratio between ion and electron is equal to 1836. The temperature ratio
between electrons and ions is equal to 10. Red lines are results from FH with $N_H=300$. Black lines are results from PIC with $N_{pcel}=102400$. The panels
 from top to bottom represent the amplitude of the modes $E_1$ to $E_4$. $E_1$ is the mode that is perturbed at $T=0$.}\label{fig:ionacoustic_time}
\end{figure}

\begin{figure}
\centering
\includegraphics[width=1\textwidth]{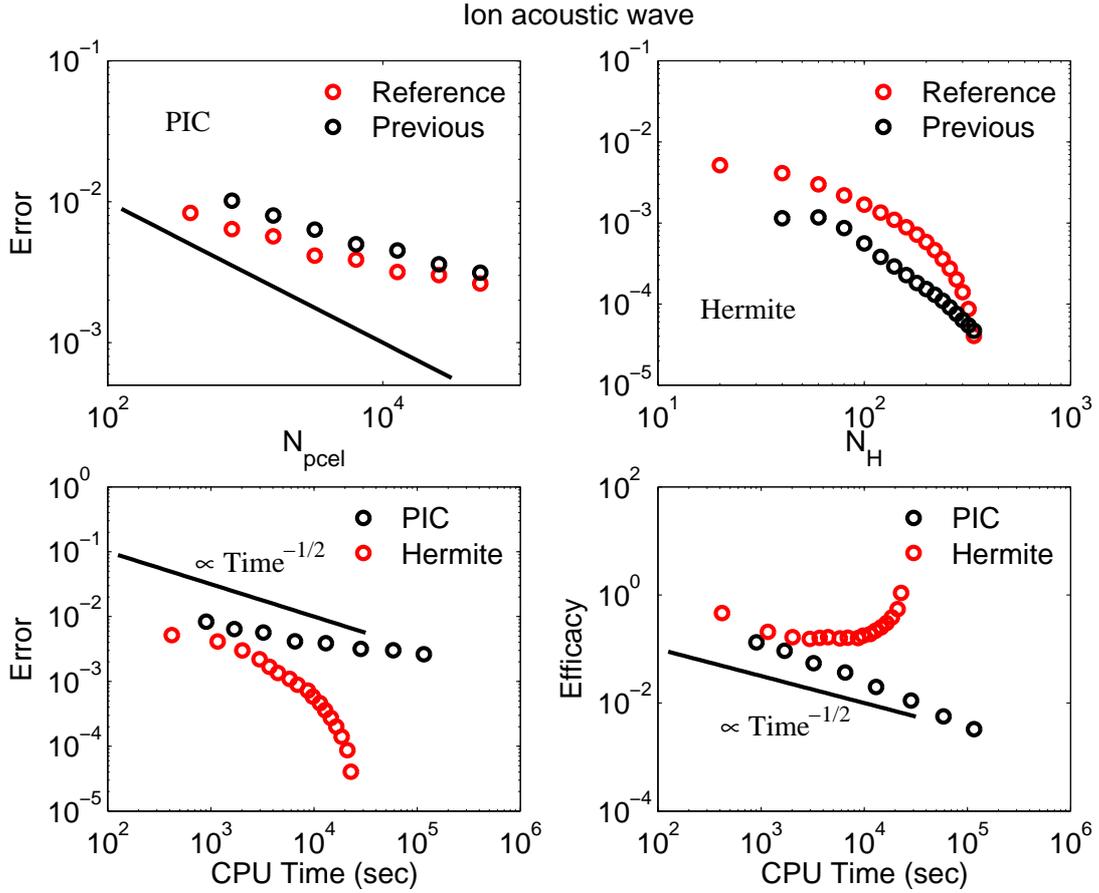}
\caption{Ion-acoustic wave with parameters $L=10$, $\alpha_e=\sqrt{2}$, $\alpha_i=0.0074\alpha_e$, $\varepsilon=0.2$, $\nu=1$. The mass ratio between ion and electron is equal to 1836. The temperature ratio
between electrons and ions is equal to 10. Top left: PIC simulation; error as a function of number of particles per cell $N_{pcel}$. Red and black circles represent the error
calculated with respect to a reference solution (with $N_{pcel}=102400$) and previous less accurate solution, respectively. The black solid line indicates the scaling $N_{pcel}^{-1/2}$.
Top right: Hermite simulation; error as a function of number of Hermite modes $N_H$. Red and black circles represent the error
calculated with respect to a reference solution (with $N_H=400$) and previous less accurate solution, respectively.
Bottom left: error as a function of CPU time (in seconds); black circles for PIC, red circles for Hermite. Bottom right: efficacy as a function of CPU time (in seconds); black circles for PIC, red circles for Hermite.
}\label{fig:ionacoustic}
\end{figure}

The ion-acoustic wave is often used as a benchmark test since it involves multi-scale physics.
In fact, it is generated by a perturbation in the ion density only, and has 
a frequency much lower than the electron plasma frequency, but it still requires accurate representation of the electron dynamics.
The parameters for this case are the following: The mass ratio between ions and electrons is equal to 1836. The ratio
between electron and ion temperature is equal to 10, $L=10$, $\alpha_e=\sqrt{2}$, $\varepsilon=0.2$. The collisionality is $\nu=1$.
Similar to the ion-acoustic wave case discussed in Part 1, the initial perturbation is large enough to drive nonlinear interactions.
Figure \ref{fig:ionacoustic_time} shows the time evolution of the amplitude of the first 4 Fourier modes.
$E_1$ is the initially excited mode, and its higher harmonics ($E_2$, $E_3$, etc.) are excited via wave-wave interactions.
The red lines are the results from FH with $N_H=300$ and the black lines are results from the PIC reference solution (with $N_{pcel}$=102400).
For the fundamental mode (top panel), there is good agreement between FH and PIC, although one can notice a certain amount of noise in the PIC results.
The first harmonic $E_2$ is quite noisy in the PIC results, but there is still a qualitative agreement between the two methods,
with the PIC solution having the correct order of magnitude and approximately the correct frequency.
However, for the PIC solution, 
harmonics higher than the second ($E_3$ and $E_4$ plotted respectively in the third and last panel) 
are at the noise ground level and completely miss the correct physical evolution.
This is an example where, although the fundamental mode has been excited at a large (nonlinear) amplitude, the PIC method is
able to capture only the largest scale fluctuations, completely missing the lower amplitude interactions.
We have described a similar shortcoming of PIC in the companion paper \citep{camporeale13a}, for a very different physical phenomenon (the plasma echo).
As we anticipated, it is not surprising that the numerical noise known to plague PIC plays
a similar role in explicit, semi-implicit and fully-implicit versions of the algorithm, even though one expects the latter
to be the most accurate of the three methods.\\
The performance/efficacy study for the ion-acoustic test is presented in Figure 
\ref{fig:ionacoustic}. 
One can see that this is a much harder test, and in general the errors are larger than in the previous examples for similar $N_{pcel}$ and $N_H$.
For this case, the PIC method recovers the theoretical scaling with $N_{pcel}$ only for the first few data points
(top-left panel). This is an indication that even the most accurate solution ($N_{pcel}=102400$) taken as a 'reference' is, 
in reality still far from convergence. The FH method is also not yet in the full spectral convergence regime.
With regards to the comparison between PIC and FH performance and efficacy, conclusions similar to previous cases hold.
Note that this case has a generally longer CPU time with respect, for instance, to the Landau damping case, simply because we
have run the simulations for many more timesteps (the ion-acoustic wave frequency is about 60 times lower than the electron plasma frequency).
The errors of PIC and FH are of similar orders of magnitude for CPU times smaller than 1000 seconds (left-bottom panel). However, the 
gap in errors between the two methods becomes as large as two orders of magnitude for CPU times of about 20,000 seconds.

\subsection{Two-stream instability}\label{twostream}
\begin{figure}
\centering
\includegraphics[width=1\textwidth]{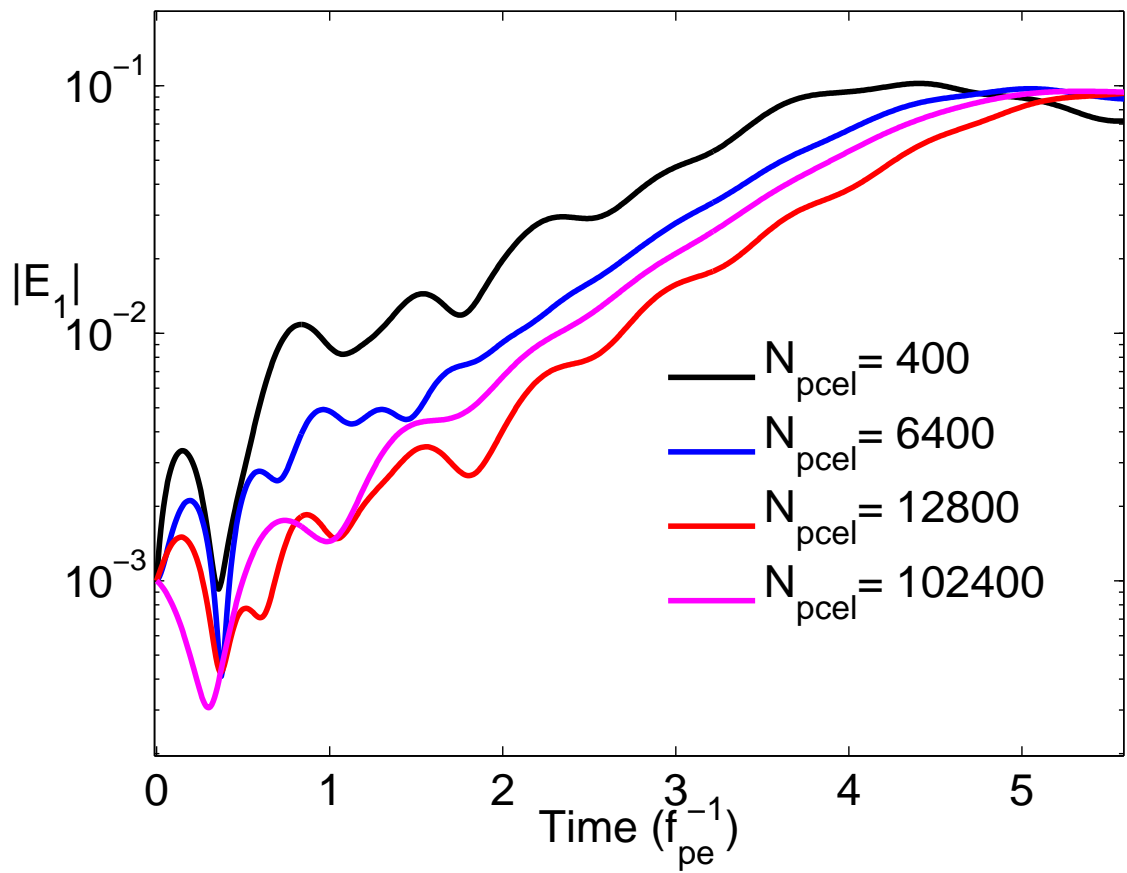}
\caption{Two-stream instability simulation with PIC. Black, blue, red, and magenta lines are for $N_{pcel}=400, 6400, 12800, 102400$, respectively.}\label{fig:twostream_time}
\end{figure}

\begin{figure}
\centering
\includegraphics[width=1\textwidth]{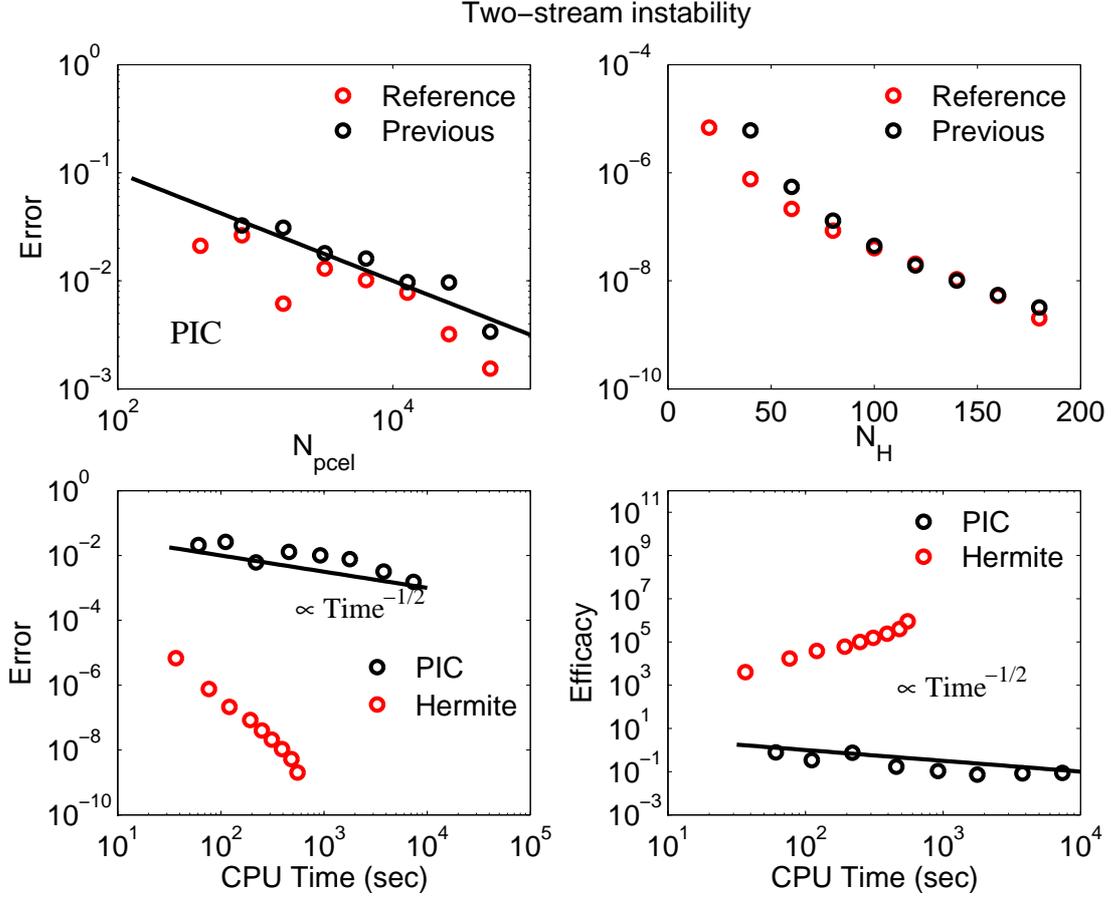}
\caption{Two-stream instability with parameters $L=4\pi$, $\alpha_e=\sqrt{2}/2$, $U=1$, $\varepsilon=0.001$, $\nu=2$. Top left: PIC simulation; error as a function of number of particles per cell $N_{pcel}$. Red and black circles represent the error
calculated with respect to a reference solution (with $N_{pcel}=102400$) and previous less accurate solution, respectively. The black solid line indicates the scaling $N_{pcel}^{-1/2}$.
Top right: Hermite simulation; error as a function of number of Hermite modes $N_H$. Red and black circles represent the error
calculated with respect to a reference solution (with $N_H=100$) and previous less accurate solution, respectively.
Bottom left: error as a function of CPU time (in seconds); black circles for PIC, red circles for Hermite. Bottom right: efficacy as a function of CPU time (in seconds); black circles for PIC, red circles for Hermite.}\label{fig:twostream}
\end{figure}

The two-stream instability is a classical textbook study often used as a benchmark for kinetic plasma codes.
It is a linear instability that is excited when the plasma consists of two populations of particles counter-streaming with a large enough relative speed.
We initialize the electron distribution function as
two drifting Maxwellians with equal temperature:
\begin{equation}
 f_e=0.5e^{\left(\frac{v-U}{\alpha_{e}}\right)}+0.5e^{\left(\frac{v+U}{\alpha_{e}}\right)},\label{drift-maxw}
\end{equation}
with $U$ the drift velocity.\\
The distribution function in Eq. (\ref{drift-maxw}) can be efficiently described in the FH method as
two distinct electron populations by setting $u_e=\pm U$ and solving Eq.(\ref{vlasov_cn}) separately for each drifting Maxwellian.
The parameters are as follows: $L=4\pi$, $\alpha_e=\sqrt{2}/2$, $U=1$, $\varepsilon=0.001$, $\nu=2$.
Figure \ref{fig:twostream_time} shows the time evolution of $E_1$ for PIC solutions
with different numbers of particles per cell (black: $N_{pcel}=400$, blue: $N_{pcel}=6400$, red: $N_{pcel}=12800$, 
magenta: $N_{pcel}=102400$). 
As already noted in Part 1, the onset of the two-stream instability is very sensitive to the initial condition.
Although the same linear growth rate is approximately recovered in all cases, the solutions for times $T<2$
are completely different.
Figure \ref{fig:twostream} shows the comparison of performance between FH and PIC, with the same format of previous Figures.
For a CPU time around 550 seconds, the FH solution is about $10^7$ times more accurate than the PIC solution.
Also, the most accurate PIC solution ($N_{pcel}=51200$) is less accurate than the least accurate FH solution.

\section{Conclusions}
This paper concludes the study initiated in the companion paper (Part 1) \citep{camporeale13a}. We have described a
spectral method to numerically solve the Vlasov-Poisson equations that describe the evolution of a collisionless plasma.
The velocity and configuration spaces are discretized by means of an Hermite and Fourier basis, respectively.
In this paper we have introduced an implicit Cranck-Nicolson discretization in time, which is charge, momentum, and energy conserving.
The simultaneous conservation of these three quantities is a very important property of this method, especially when compared with PIC methods.
Indeed, no PIC code is able to simultaneously conserve momentum and energy and thus require monitoring of the non-conserved quantity.\\
We have compared the performance of the implicit FH method with the recently proposed implicit PIC code for the case of one-dimensional, electrostatic simulations.
The comparison results for Langmuir wave, Landau damping, ion-acoustic wave, and two-stream instability are summarized in Figures \ref{fig:langmuir}, \ref{fig:landau}, \ref{fig:ionacoustic},
\ref{fig:twostream}, respectively.
The two metrics that we have considered in order to fairly assess which method is computationally more advantageous are errors and efficacy as a function of CPU time.
Notice that
\citet{chen11,chen13} have presented performance comparisons between the fully-implicit and the explicit PIC (which is forced to use smaller timesteps), showing that the CPU time for the explicit PIC
is tens or hundreds of time larger than the corresponding CPU time for implicit PIC. 
This comparison, however, does not take into account the accuracy of the solution, 
which is different in the two methods (the explicit simulation is usually more accurate since it uses smaller timestep and/or cell size, 
but might return a much smaller error than what acceptable in practice. See for instance \citep{camporeale13} for a concrete example where some of these considerations are discussed), 
but can be crucial since one would want the faster algorithm for a given accuracy. Therefore, here and in Part 1 we have opted for a different metric, 
where we evaluate the efficacy of a simulation, computed as the inverse of the product of its error (relative to an accurate 'reference' solution) and the CPU time. 
The efficacy is a measure of the cost-effectiveness of a method or, in other words, how much a larger CPU time is paid off in terms of better accuracy. \\
The trends shown are consistent with the results presented in Part 1.
In particular, although each simulation is quantitatively different, they all share the same conclusion that the FH method is orders of magnitude more efficient than PIC, at least for the examples considered.
Similar to the plasma echo example discussed in Part 1, we have shown, for the test case of the ion-acoustic wave, that the PIC code is unable to 
correctly capture the higher-order harmonics of the excited dominant mode (excited at nonlinear amplitude), making  any analysis of wave-wave coupling and energy transfer impossible.
This is an important problem that can impact PIC applications in areas such as plasma turbulence (see, e.g. \citep{haynes13}), suggesting that denoising techniques are mandatory for PIC there.
In conclusion, we believe that the spectral expansion of the velocity space such as in the 
FH method is a very promising path that should be more systematically explored for accurate and 
fast simulations of kinetic collisionless plasmas, especially in light of the upcoming exascale era and the cost of high-performance computer 
simulations that will require an efficient use of the available resources. 
As discussed in Part 1, further developments of the FH method in terms of optimization of the Hermite basis are critical for 
multidimensional applications of this technique, and we hope that this paper can stimulate some research in this area.

\section{Acknowledgement}
The authors would like to thank L. Chacon, S. Markidis, V. Roytershteyn, X. Tang for useful conversations.
This work was funded by the Laboratory Directed Research
and Development program (LDRD), U.S. Department of Energy Office of
Science, Office of Fusion Energy Sciences, under the auspices of the
National Nuclear Security Administration of the U.S. Department of
Energy by Los Alamos National Laboratory, operated by Los Alamos
National Security LLC under contract DE-AC52-06NA25396.





\bibliographystyle{plainnat}






\newpage
\begin{figure}
\centering
\includegraphics[width=1\textwidth]{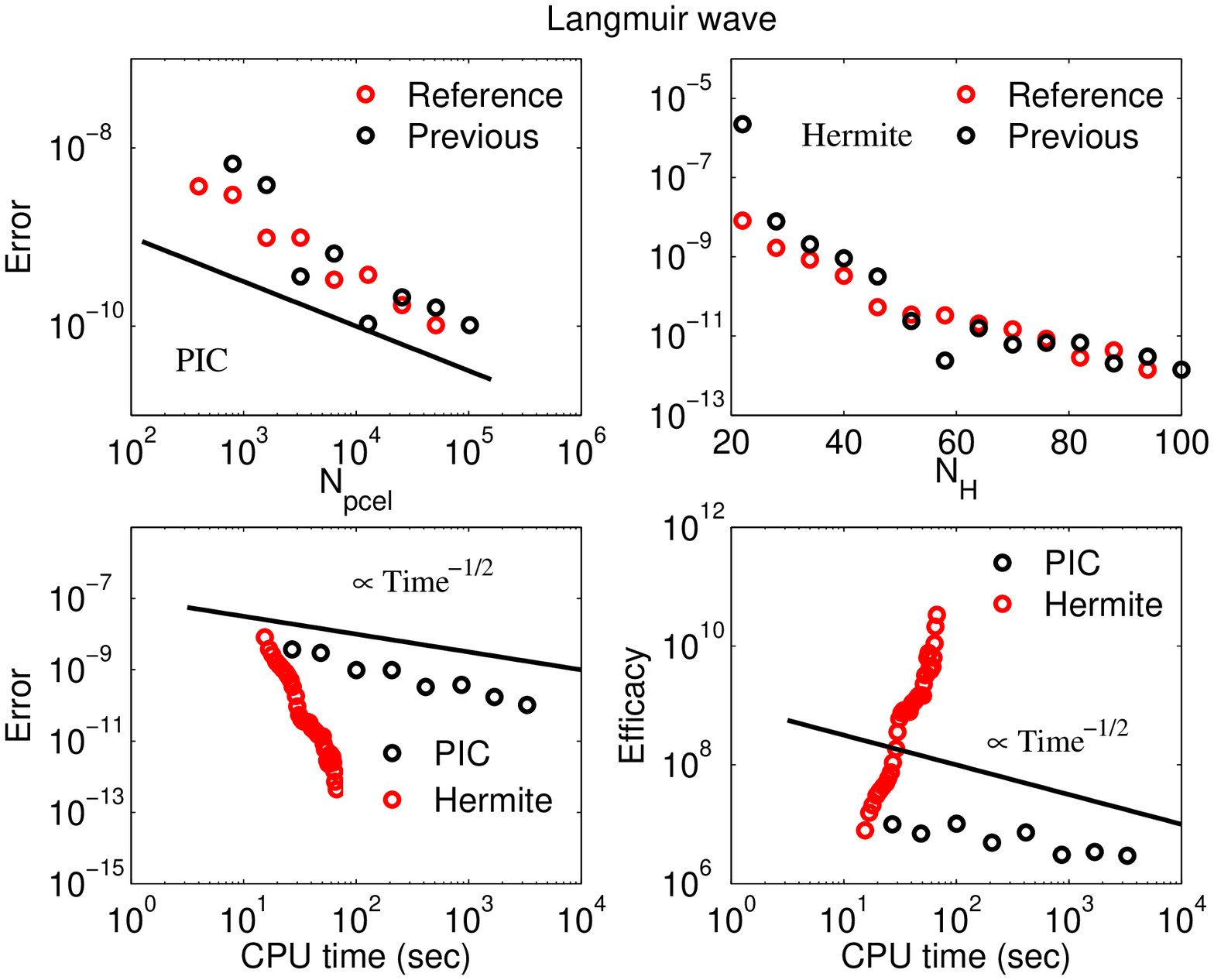}
\caption{Langmuir wave with parameters $L=2\pi$, $\alpha_e=0.1\sqrt{2}$, $\varepsilon=0.01$, $\nu=0$. Top left: PIC simulation; error as a function of number of particles per cell $N_{pcel}$. Red and black circles represent the error
calculated with respect to a reference solution (with $N_{pcel}=102400$) and previous less accurate solution, respectively. The black solid line indicates the scaling $N_{pcel}^{-1/2}$.
Top right: Hermite simulation; error as a function of number of Hermite modes $N_H$. Red and black circles represent the error
calculated with respect to a reference solution (with $N_H=100$) and previous less accurate solution, respectively.
Bottom left: error as a function of CPU time (in seconds); black circles for PIC, red circles for Hermite. Bottom right: efficacy as a function of CPU time (in seconds); black circles for PIC, red circles for Hermite.
}\label{fig:langmuir}
\end{figure}

\end{document}